# Accelerating Diffusion Models for Generative AI Applications with Silicon Photonics

Tharini Suresh
Electrical and Computer Engineering,
Colorado State University,
Fort Collins, USA
Tharini.Suresh@colostate.edu

Salma Afifi
Electrical and Computer Engineering,
Colorado State University,
Fort Collins, USA
Salma.Afifi@colostate.edu

Sudeep Pasricha
Electrical and Computer Engineering,
Colorado State University,
Fort Collins, USA
Sudeep@colostate.edu

*Abstract*—Diffusion models have revolutionized generative AI, with their inherent capacity to generate highly realistic state-of-the-art synthetic data. However, these models employ an iterative denoising process over computationally intensive layers such as UNets and attention mechanisms. This results in high inference energy on conventional electronic platforms, and thus, there is an emerging need to accelerate these models in a sustainable manner. To address this challenge, we present a novel silicon photonics-based accelerator for diffusion models. Experimental evaluations demonstrate that our photonic accelerator achieves at least 3× better energy efficiency and 5.5× throughput improvement compared to state-of-the-art diffusion model accelerators.

## I. INTRODUCTION

Diffusion models (DMs) have emerged as a powerful class of generative AI models for image synthesis, video creation, and text-to-image generation [1]. DMs are being explored across a wide range of fields such as electronic design automation (EDA) [2], drug discovery by generating novel molecular structures [3], and in medical imaging by generating synthetic scans to enhance datasets for training medical diagnostic models.

DMs work by progressively adding noise to data over a series of timesteps in a forward process and learning to reverse this process step-by-step to recover the original data. Examples of prominent DMs include Denoising Diffusion Probabilistic Models (DDPM) [1], Latent Diffusion Models (LDMs) [4] and Stable Diffusion Models (SDMs) [5]. While DDPMs operate directly in the data space, LDMs and SDMs perform diffusion in a compressed latent space.

Despite their ability to generate high quality synthetic data, the iterative denoising process inherent in DMs requires a significant number of steps, each demanding intensive computation. These repeated steps, necessary for generating high-quality data, lead to high latency, making it difficult to apply DMs efficiently on resource-constrained devices or for time-sensitive tasks. Moreover, the need for precise noise estimation at each step adds further complexity, often resulting in performance bottlenecks on today's electronic hardware platforms such as GPUs and various accelerators.

The limitations of conventional electronic platforms in supporting diffusion models are further exacerbated in the post-Moore's law era [6], where continued transistor scaling no longer offers the significant performance and energy gains that it did in the past. Further, data transmission over metallic interconnects introduces significant bandwidth and power bottlenecks, while fabrication costs continue to rise. In today's rapidly evolving AI landscape, there is also a growing consensus on developing environmentally sustainable hardware platforms for AI that minimize energy overheads. Neural network accelerators that leverage silicon photonics [7] are a promising solution to overcome the limitations of conventional electronic architectures. Photonic integrated circuits (PICs) enable energy-efficient computation and data movement in the optical domain, and are compatible with CMOS fabrication processes. Furthermore, matrix-vector multiplications, an integral component of neural network workloads, can be performed using silicon photonic devices.

In this paper, we present an overview of the first silicon photonics-based accelerator that is capable of accelerating inference of a broad family of DMs. Our proposed accelerator demonstrates significant enhancements in both throughput and energy efficiency when compared to the state-of-the-art DM hardware accelerators.

## II. RELATED WORK

Silicon photonics has been shown to efficiently accelerate deep neural networks. Several efforts have demonstrated that silicon photonics can be used to accelerate convolution neural networks (CNNs), binary neural networks (BNNs), recurrent neural networks (RNNs), large language models (LLMs), graph neural networks (GNNs), and generative adversarial networks (GANs) [8]-[17]. In addition to targeting specific neural network types, several works have proposed silicon photonic accelerators with optimized co-design support. For example, in [18], an optical accelerator for CNNs with support for homogeneous and heterogeneous model quantization was proposed. Some recent efforts have also focused on accelerating CNNs with photonic-based processing in-memory (PIM). In [19], an Optically-addressed Phase Change Memory (OPCM)-based PIM system was proposed to mitigate the programming overhead during CNN inference in OPCM arrays. In [20], an OPCM-based main memory was proposed, that enables in-place CNN inference with high parallelism and energy efficiency. Despite these promising efforts, leveraging silicon photonics to accelerate diffusion models (DMs) has not yet been addressed, due to the challenging computational demands of these generative models. In this work, we present the first silicon photonic-based generative DM accelerator.

Prior efforts to accelerate DMs have focused on either reducing the total number of timesteps or reducing the time taken per step. For instance, [21] proposed a method to reduce the model size per timestep by caching high level features across adjacent timesteps, thereby reducing average latency per timestep for DDPM, LDM and SDM models. However, the high memory demands of this approach limit scalability. An FPGA-based SDM accelerator [22] proposed customized compute units for matrix multiplication, data layout transformations, and vector/scalar operations. The design improves energy efficiency over GPU and CPU baselines, but suffers from high inference latency. Another FPGA-based solution [23] introduced a hybrid systolic array architecture that supports both convolution and attention operators through efficient pipelining for end-to-end SDM acceleration.

## III. BACKGROUND: DIFFUSION MODELS

### A. Diffusion Models

DMs operate via a two-stage process, as illustrated in Figure 1. In the forward process, Gaussian noise is added progressively to a data sample $x_0$ over $T$ timesteps. At each step $t$, a noisier sample $x_t$ is generated as:

$$x_t = \sqrt{1-\beta_t}\, x_{t-1} + \sqrt{\beta_t}\, \epsilon, \quad \epsilon \sim \mathcal{N}(0, I) \quad (1)$$

where $\beta_t$ controls the noise level at each step, and $\epsilon$ is standard Gaussian noise. This process defines a Markov chain that transforms $x_0$ into pure noise, modeled by the conditional distribution $q(x_{1:T}|x_0)$.

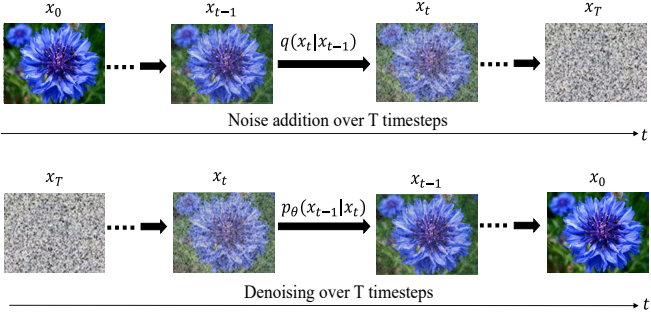

Figure 1: Diffusion model operation.

In the reverse diffusion process, the model performs iterative denoising over $T$ timesteps. The denoising process is modeled by the learned distribution $p_\theta(x_{t-1}|x_t)$ which predicts a less noisy sample $x_{t-1}$. At each timestep $t$, a neural network (typically a UNet) predicts the mean $\mu_\theta(x_t, t)$ of this distribution, and generates a less noisy sample as:

$$x_{t-1} = \mu_\theta(x_t, t) + \sigma_t z, \quad z \sim \mathcal{N}(0, I) \quad (2)$$

where $\sigma_t$ is a fixed parameter that controls how much noise is added back at each step, and $z$ is random Gaussian noise. During inference, the model starts from random noise and iteratively applies the reverse process to generate a high-quality sample.

A UNet in DMs consists of stacked downsampling (encoder) and upsampling (decoder) blocks with skip connections to combine low-level and high-level features for precise image reconstruction. Multi-head attention (MHA) layers, each containing multiple self-attention heads, are integrated into the U-Net within DMs. Each self-attention head calculates attention scores using query ($Q$), key ($K$), and value ($V$) vectors, which are derived by multiplying the input feature map $X$ by the query, key, and value weight matrices: $W_Q$, $W_K$ and $W_V$, respectively. The attention output is then computed using scaled dot-product attention:

$$Head(X) = Attention(Q, K, V) = softmax\left(\frac{QK^T}{\sqrt{d_k}}\right). V \quad (3)$$

where $X$ is the input matrix and $d_k$ is the dimension of $Q$ and $K$. The output of all the self-attention heads are then concatenated and passed through a linear layer, to obtain the final MHA output.

The log-sum-exp approach can be used to express the softmax function used in self-attention heads as:

$$softmax(\gamma_i - \gamma_{max}) = \frac{\exp(\gamma_i - \gamma_{max})}{\Sigma_{j=1}^D \exp(\gamma_j - \gamma_{max})}$$

$$= \exp\left(\gamma_i - \gamma_{max} - \ln\left(\Sigma_{j=1}^D \exp(\gamma_j - \gamma_{max})\right)\right) \quad (4)$$

Using this method, the computationally intensive softmax can be broken down into four main operations: 1) identifying $\gamma_{max}$; 2) calculating $\ln\left(\Sigma_{j=1}^D \exp(\gamma_j - \gamma_{max})\right)$; 3) subtracting the $ln$ output from $(\gamma_j - \gamma_{max})$; and 4) $exp$ of the final value. This breakdown helps to better exploit the inherent parallelism in silicon photonics.

Lastly, DM variants differ in their computational demands, which complicates accelerator design. DDPMs, which operate in the pixel space, involve higher-dimensional feature maps and are dominated by convolutional operations. LDMs reduce spatial complexity by operating on compressed representations but require additional encoding and decoding blocks. SDMs further extend LDMs by incorporating additional attention layers within the UNet, thus increasing the relative importance of attention-heavy operations during inference. An efficient DM accelerator must be able to flexibly adapt its computational pipeline to the specific needs across the various diffusion model variants.

### B. Silicon Photonics Components for DM Acceleration

Optical neural network accelerators can be broadly classified into coherent or non-coherent implementations [31]. Coherent architectures use a single wavelength, where parameters are imprinted onto the optical signal's phase. This enables Multiply and Accumulate (MAC) operations through phase modulation. In contrast, non-coherent architectures leverage multiple wavelengths, imprinting parameters onto the amplitude of the optical signal. This enables parallel MAC operations via intensity modulation across different wavelengths, greatly enhancing throughput. For this reason, in our photonic DM accelerator, we utilize this approach.

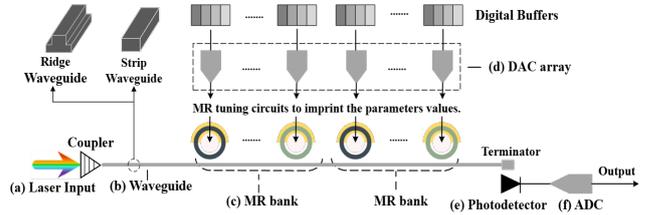

Figure 2: Fundamental silicon photonic components for DM acceleration.

Figure 2 provides an overview of the fundamental devices and circuits required for non-coherent computing with silicon photonics. The following are the main components needed:

*1) Lasers* are the source of optical signals essential for both computation and communication. These lasers can be integrated on-chip, such as vertical cavity surface-emitting lasers (VCSELs), which offer better integration and reduced propagation losses. Alternatively, off-chip lasers can be used to achieve higher optical power and efficiency, though they are associated with significant losses when coupling the optical signals onto on-chip waveguides.

*2) Waveguides* transmit the optical signals generated by the laser source and are typically fabricated using high-refractive-index contrast materials, such as a silicon (Si) core surrounded by a silicon dioxide ($SiO_2$) cladding, to realize total internal reflection that confines the light within the core region. By utilizing Wavelength Division Multiplexing (WDM), a single waveguide can carry multiple wavelengths simultaneously, enabling parallel multiply-accumulate (MAC) operations (one operation per wavelength).

*3) Microring Resonators (MRs)* are optical modulators that are capable of performing MAC operations by selectively modulating input signals onto their resonant wavelengths. Each MR can be tuned to a specific wavelength ($\lambda_{MR}$), defined as $\lambda_{MR} = \frac{2\pi R}{m} n_{eff}$, where $R$ is the fabricated MR's radius, $m$ is the order of resonance, and $n_{eff}$ is the effective index of the device. MRs are used to imprint input activations and weights onto optical signals for the matrix multiplications in DMs.

*4) Photodetectors (PDs)* convert the processed optical signals into electrical signals. They need to exhibit high sensitivity to accurately detect weak optical inputs and compensate for optical losses (e.g., propagation, bending, through) incurred along the optical signal path.

*5) Tuning circuits* are required to carefully control the effective index $n_{eff}$ of active MR devices. These circuits induce a resonant shift $\Delta\lambda_{MR}$ in the MR's operating wavelength, which is required for accurately modulating electronic data onto optical signals. These fine adjustments to the resonant wavelength coupling are essential for achieving error-free modulation during optical computations.

*6) Digital-to-Analog Converters (DACs)* and *Analog-to-Digital Converters (ADCs)* are essential for appropriately tuning the MRs and converting optical signals to the digital domain for intermediate processing, respectively. However, both of these are high latency and power-hungry components, contributing significantly to the energy overhead of silicon photonic systems.

## C. MAC Operations in the Optical Domain

Non-coherent optical computations are performed by modifying the resonant wavelength of an MR device ($\Delta\lambda_{MR}$), resulting in a predictable change in the amplitude of the resonant optical signal's wavelength. Vector dot product of two activations ($a1$, $a2$) and two weights ($w1$, $w2$) is achieved by utilizing optical signals of different wavelengths multiplexed into a single waveguide. An initial MR bank modulates the optical signals with activation values $a1$ and $a2$, while a subsequent MR bank along the optical signal path imprints the corresponding weight values $w1$ and $w2$ onto the same optical signals, resulting in multiplication operations. These modulated signals are then passed through a photodetector (PD) to accumulate the result of the dot product ($a1w1 + a2w2$). Summation of two signals, $a1$ and $a2$ can be performed using coherent photonic summation, where the signals are assigned the same wavelengths and share a physical waveguide path that induces an aggregation of their individual optical signal intensities. As multiplications and accumulations (MAC operations) dominate the runtime of neural networks [31], light domain implementation of these operations can significantly improve overall throughput and energy efficiency.

## IV. DIFFLIGHT HARDWARE ACCELERATOR

An overview of our proposed diffusion model accelerator is shown in Figure 3. The architecture consists of a Residual unit and a multi-head attention (MHA) unit. The Residual unit consists of $Y$ convolution and normalization blocks, and an activation block. The MHA unit consists of $H$ attention head blocks, and a linear and add block. An integrated electronic control unit (ECU) performs tasks such as interfacing with electronic memory, buffering intermediate results, and mapping matrices to the photonic domain. Each dense and convolution block utilizes a single VCSEL array to supply the necessary optical signals across the rows in the MR bank arrays. This VCSEL reuse strategy not only minimizes the power consumption associated with laser sources but also reduces the potential for inter-channel crosstalk [32], ensuring the integrity of optical signals within the system.

In the rest of this section, we discuss the key building block components of our proposed photonic DM accelerator.

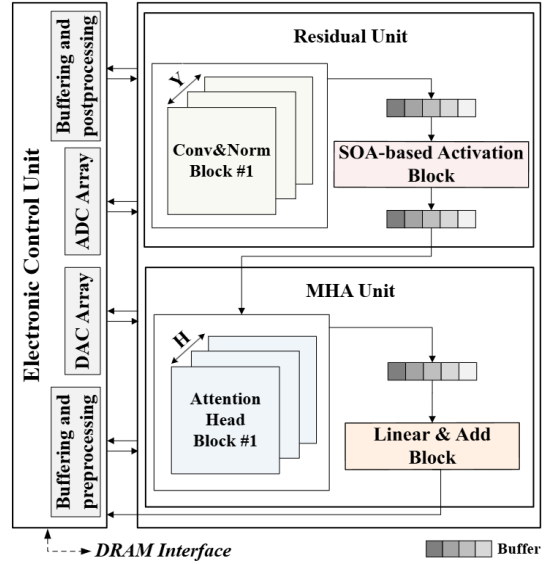

Figure 3: Overview of our proposed photonic accelerator architecture.

## A. MR tuning circuit design

Precise tuning of MR devices is essential for efficient optical computation. A hybrid tuning circuit that leverages electro-optic (EO) [24] and thermo-optic (TO) tuning [25] methods to adjust the MR resonant wavelength ($\Delta\lambda_{MR}$) is used in our accelerator. EO tuning, with low power consumption (≈4 μW/nm) and fast response times (≈ns range), is used for small wavelength adjustments. This is supplemented by TO tuning that offers a larger tunability range, but with the drawback of higher latency (≈μs range) and power consumption (≈27 mW/FSR). Thus the fast EO tuning is used by default while TO tuning is initiated sporadically as needed to ensure accuracy despite significant environmental changes (e.g., increase in chip temperature to elevated levels). To further optimize power and reduce thermal crosstalk (induced by tuning), we employ the Thermal Eigenmode Decomposition (TED) method [26]. TED ensures efficient TO tuning by minimizing interference between neighboring MRs, reducing overall power consumption. The hybrid tuning approach allows the accelerator to maintain high-speed and low-power operation across different model workloads.

## B. Architecture Design

### 1) Convolution and Normalization Block

Each block is implemented optically using two MR bank arrays, with dimensions $K \times N$, where $K$ represents the number of rows (each row comprising both a positive and a negative value waveguide), and $N$ represents the number of columns, as shown in Figure 4. The first MR bank in each block is responsible for imprinting the input activations, while the second bank imprints the weight values onto the optical signals. The modulated optical signals are then detected by balanced photodetectors (BPDs) at the end of each block, producing a final accumulated analog value that represents the weighted sum of the inputs. BPDs are specialized PDs featuring two distinct arms connected to the same row—one for positive signal polarities and the other for negative ones. This design enables them to handle both positive and negative parameter values by measuring the absolute difference between the two signals. The BPD calculates the net difference signal by subtracting the output of the negative arm from the positive arm.

Broadband MRs are used to implement normalization, including Group Normalization, which is commonly used in DMs due to its stability across varying batch sizes. The normalization parameters, which are updated during inference, can be used to tune the broadband MRs for real-time adjustment. The broadband MRs can also be bypassed, when normalization is not required.

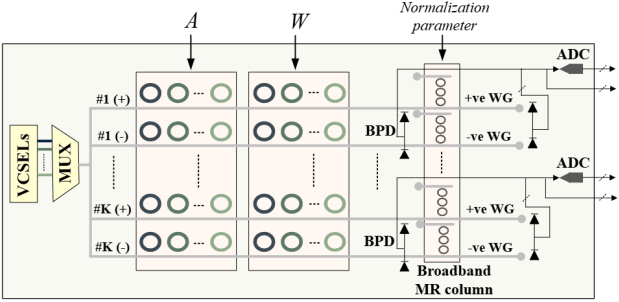

Figure 4: Convolution and normalization block consisting of two MR bank arrays and a broadband MR bank.

*2) Activation Block*

Previous works have demonstrated that semiconductor-optical-amplifiers (SOAs) can be leveraged to optically implement non-linear functions such as sigmoid [27]. We propose an optical implementation of the swish activation function, which is commonly used in DMs, and defined as:

$$f(x) = x \cdot sigmoid(x) \quad (5)$$

As shown in Figure 5, the input $x$ is used to drive a VCSEL, and the resulting optical signal is sent to the SOA-based sigmoid block. The analog output of the sigmoid block is detected by a PD and used to tune an MR on the next waveguide. The multiplication between $x$ and $sigmoid(x)$ is implemented using an MR, as shown. To add the residual connection that follows activation layers, coherent photonic summation (discussed earlier) is used.

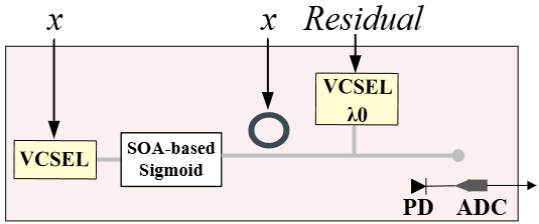

Figure 5: SOA-based implementation of swish activation function.

*3) Attention Head Block*

Each block is implemented optically using seven MR banks as shown in Figure 6, with the softmax function being implemented in the ECU. Using MatMul decomposition, $Q.K^T$ can be expressed as:

$$Q.K^T = Q.(X.W_K)^T = (Q.W_K^T).X^T \quad (6)$$

The four MR banks in the upper part of the attention head block are used to realize this operation, with each MR bank array having dimensions $M \times L$, where $M$ represents the number of rows, each comprising both a positive and a negative waveguide, and $L$ represents the number of columns. The first two MR banks generate $Q$, followed by two MR banks which modulate $W_K^T/\sqrt{d_k}$ and $X^T$ values consecutively. By introducing the scaling factor $\sqrt{d_k}$ as part of the weight matrices, we reduce the scaling overhead. The final partial sums are accumulated using BPDs. The results are then converted to the digital domain, using an analog-to-digital converter (ADC) to undergo softmax computation.

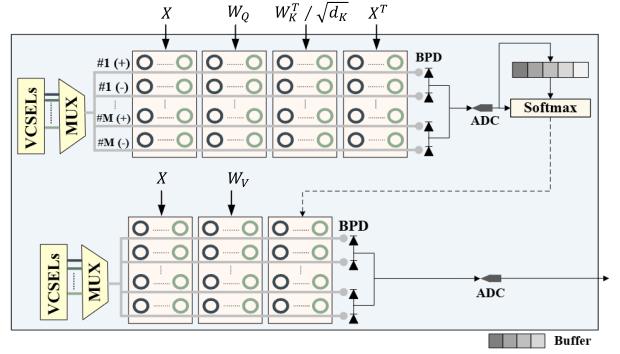

Figure 6: Attention head block consisting of seven MR bank arrays.

To enable pipelined execution of softmax, as the attention score values are being generated and digitized via the ADC, they are immediately buffered and fed into a comparator circuit to update and track $\gamma_{max}$, enabling concurrent execution of softmax sub-operations. Simultaneously, in the lower part of the attention head block, we generate $V$ using two MR banks of dimensions $M \times N$. Further, in the ECU, a subtractor circuit is utilized to compute $\gamma_j - \gamma_{max}$, and lookup tables (LUTs) are leveraged for $ln$ and $exp$ to obtain the final softmax values. The attention matrix $Attn$, output from the softmax computation, is then modulated into the third MR bank to obtain the final attention head output, which is detected by BPDs. The outputs of all the attention head blocks are buffered and concatenated, and the final value is then passed to the linear and add block.

*4) Linear and Add block*

The MHA unit also consists of a single linear and block. As shown in Figure 7, it has a linear path comprising of two MR bank arrays, one for input activations and the other for weights, each with dimensions $M \times L$. The modulated optical signals are detected by BPDs. Next, in the add path, the final accumulated output is used to drive a VCSEL operating at wavelength $\lambda_o$. Another VCSEL, also functioning at the same wavelength $\lambda_o$ is used to generate an optical signal with the residual parameter imprinted onto it. The two values undergo coherent summation after that and the final analog value is detected by a PD.

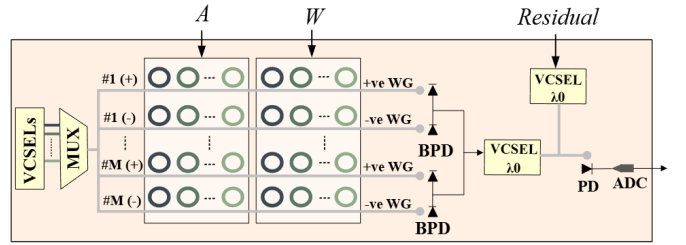

Figure 7: Linear and Add block

## C. Dataflow and scheduling optimizations

The decoder blocks in the DM UNet consist of transposed convolutions, which expand the input feature map using zero-insertion, before applying a convolutional kernel. The kernel slides over the expanded input to compute the output feature map. The dot product operations involving zeroes are inefficient, and lead to resource underutilization. We adopt a sparsity-aware dataflow that identifies and eliminates these inefficient operations. In this approach, the all-zero columns in the flattened input feature map and corresponding elements in the flattened kernel are identified, and corresponding kernel elements are then eliminated, resulting in a lower-dimension dot-product operation. Pipelining at different levels of granularity – inter-block and intra-block operations – is also introduced, to achieve throughput improvement and further reduction in overall latency. Lastly, we adopt a DAC-sharing strategy, where each pair of columns in an MR bank array shares a single set of DACs in the residual and MHA units. Though this approach increases the time taken to tune all the MRs, it results in significant energy savings.

## V. EXPERIMENTS AND RESULTS

We performed detailed analysis of our proposed DM accelerator architecture (called DiffLight) across multiple diffusion model variants, as shown in Table I. We applied the industry standard W8A8 quantization algorithm [28] to all diffusion models. The inception score (IS) metric [29] was used to assess the quality of the output generated by the DMs. The datasets used to evaluate the models and the change in IS after 8-bit quantization are also shown in Table I. PyTorch 2.4.1 was used to train and evaluate the models.

TABLE I: EVALUATED DMs, DATASETS, PARAMETERS, AND IS

| Model | Dataset | Parameters | IS reduction after 8-bit quantization |
|---|---|---|---|
| DDPM | CIFAR-10 | 61.9M | 0.44 % |
| LDM 1 | LSUN-Churches | 294.96M | 0.43 % |
| LDM 2 | LSUN-Beds | 274.05M | 5.26 % |
| Stable Diffusion | sd-v1-4 | 859.52M | 6.66 % |

TABLE II: OPTOELECTRONIC DEVICE PARAMETERS

| Devices | Latency | Power |
|---|---|---|
| EO Tuning | 20 ns | 4 µW |
| TO Tuning | 4 µs | 27.5 mW/FSR |
| VCSEL | 0.07 ns | 1.3 mW |
| Photodetector | 5.8 ps | 2.8 mW |
| SOA | 0.3 ns | 2.2 mW |
| DAC (8-bit) | 0.29 ns | 3 mW |
| ADC (8-bit) | 0.82 ns | 3.1 mW |
| Comparator | 623.7 ps | 0.055 mW |
| Subtractor | 719.95 ps | 0.0028 mW |
| LUT | 222.5 ps | 4.21 mW |

For estimating the performance and energy costs of accelerating each DM, we developed a simulator in Python with the optoelectronic components discussed in Section IV accurately modeled, with latency and power values derived from various fabricated optoelectronic devices [31]. The optoelectronic devices and electronic circuit latencies and power characteristics considered are shown in Table II. The electronic circuits were synthesized using Cadence Genus while the performance and energy estimates of the LUTs and buffers utilized in our architecture were obtained using CACTI [30]. Factors contributing to photonic signal losses, such as waveguide propagation ($1\,dB/cm$), splitter ($0.13\,dB$), MR through ($0.02\,dB$) and MR modulation ($0.72\,dB$) losses are taken into account when determining appropriate laser power.

Minimizing errors during optical operations is essential to ensure reliable performance. Our photonic device-level analysis based on the Lumerical FDTD, CHARGE, MODE, and INTERCONNECT tools shows that a waveguide can accommodate up to 36 MRs for non-coherent operation, while still maintaining error-free performance and minimizing crosstalk between optical wavelengths. Our optical accelerator design adheres to this requirement.

Lastly, an in-depth design space exploration was performed to determine the number of key architectural components that offer the best throughput (in terms of GOPS or giga operations per second) and energy efficiency (in terms of EPB or energy per bit). The exploration yielded the most optimal (maximum GOPS/EPB) DiffLight configuration for architectural parameters $[Y, N, K, H, L, M]$ discussed in Section IV as $[4, 12, 3, 6, 6, 3]$. This configuration offers the best balance between performance and power efficiency.

### A. Orchestration and scheduling optimization analysis

We conducted a sensitivity analysis to assess the impact of the proposed dataflow and scheduling optimizations. The energy results shown in Figure 8 are normalized to the baseline configuration, which does not incorporate any of the optimizations. The sparse computation dataflow (S/W Optimized), pipelining, DAC sharing, and a combination of all three (S/W Optimized + Pipelined + DAC Sharing) were explored in this analysis. On average, the combined optimizations of S/W Optimized, Pipelined, and DAC Sharing result in a 3× reduction in normalized energy consumption across all models vs. the baseline. We therefore used this DiffLight configuration in subsequent comparisons.

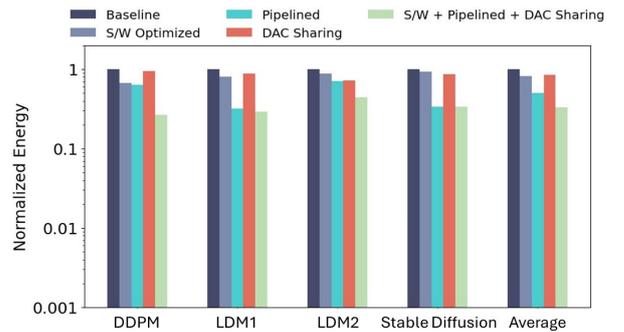

Figure 8: Energy improvements with dataflow and scheduling optimizations.

### B. Comparison with State-of-the-Art Accelerators

We compared DiffLight with multiple computing platforms and state-of-the-art DM accelerators: Nvidia RTX 4070 GPU, Intel Xeon E5-2676 v3 CPU, DeepCache [21], FPGA-based SDM accelerators FPGA_Acc1 [22] and FPGA_Acc2 [23], and photonic accelerator PACE [10]. Although existing general-purpose photonic accelerators demonstrate energy-efficient matrix vector multiplications, they are not tailored for the dataflow of diffusion models and cannot support DM-specific layers. We used our simulator to estimate GOPS and EPB for the four DM variants.

Figure 9 shows the GOPS throughput comparison. On average, DiffLight achieves 59.5×, 51.89×, 192×, 572×, 94×, and 5.5× improvement compared to CPU, GPU, DeepCache, FPGA_Acc1, FPGA_Acc2, and PACE, respectively. The high throughput gains can be attributed to DiffLight's ability to perform matrix-vector multiplication operations in the photonic domain, as well as extensive pipelining of the operations in the architecture.

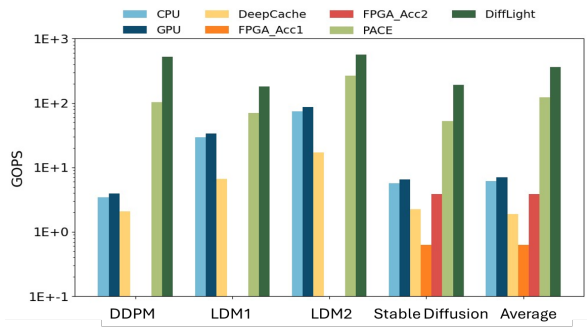

Figure 9: GOPS comparison across different diffusion models

Figure 10 shows the EPB comparison across all the DMs. On average, DiffLight exhibits 32.9×, 94.18×, 376×, 67×, 3×, and 4.51× lower EPB than CPU, GPU, DeepCache, FPGA_Acc1, FPGA_Acc2, and PACE, respectively. These significant energy improvements can be attributed to DiffLight's efficient sparsity-aware dataflow as well as DAC sharing technique, that helps eliminate unnecessary operations and improves hardware utilization.

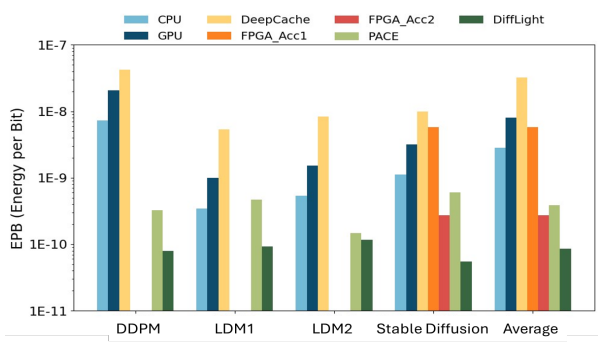

Figure 10: EPB comparison across different diffusion models

VI. CONCLUSION

In this paper, we presented the first silicon photonic accelerator for generative diffusion models. Our analysis demonstrates that our accelerator exhibits at least 5.5× higher GOPS and 3× lower EPB, compared to state-of-the-art diffusion model accelerators. These results showcase the potential of DiffLight to offer sustainable inference acceleration for diffusion models. Potential future directions for this work can involve mitigating fabrication process variations to further improve reliability [33], addressing security vulnerabilities during optical computing [34], improved dynamic optical channel sharing [35], efficient laser power management [36], and exploring in-memory optical computing [37].

ACKNOWLEDGEMENTS

This research was supported in part by National Science Foundation grant CCF-2450615.